\documentclass[showpacs, prd, twocolumn]{revtex4}
\usepackage{amsmath}
\usepackage{amssymb}
\usepackage{color}
\usepackage{oldgerm}

\usepackage{enumerate}

 \allowdisplaybreaks[1]
 
\begin{document}
 

\title{Brane Induced Gravity: From a No--Go to a No--Ghost Theorem}

\author{Felix Berkhahn}
\email{felix.berkhahn@physik.lmu.de}
\affiliation{Arnold Sommerfeld Center, Ludwig--Maximilians Universit\"at,\\
Theresienstr.~37, 80333 M{\"u}nchen, Germany}
\affiliation{Excellence Cluster Universe, Boltzmannstr.~2, 85748 Garching, Germany}

\author{Stefan Hofmann}
\email{stefan.hofmann@physik.lmu.de}
\affiliation{Arnold Sommerfeld Center, Ludwig--Maximilians Universit\"at,\\
Theresienstr.~37, 80333 M{\"u}nchen, Germany}
\affiliation{Excellence Cluster Universe, Boltzmannstr.~2, 85748 Garching, Germany}

\author{Florian Niedermann}
\email{florian.niedermann@physik.lmu.de}
\affiliation{Arnold Sommerfeld Center, Ludwig--Maximilians Universit\"at,\\
Theresienstr.~37, 80333 M{\"u}nchen, Germany}
\affiliation{Excellence Cluster Universe, Boltzmannstr.~2, 85748 Garching, Germany}

\date{\today}

\begin{abstract}
Numerous claims in the literature suggest that gravity induced on a higher 
co--dimensional surface violates unitarity in the weak coupling regime. 
However, it remained unclear, why a conserved source localized on this surface
and giving rise to an induced gravity term at low energies would absorb and emit 
the associated ghost, given a consistent source--free theory. 
In this article it is shown that the appearance of the induced Einstein Hilbert term 
does not threaten the unitarity of the theory. It is shown that the would--be ghost 
highlighted in previous works is non--dynamical and therefore not associated with a state in the Hilbert space. 
The physics arguments behind this statement are presented in a semi--covariant
language, but the detailed proof is given using Dirac's constraint analysis. The Hamiltonian 
on the constraint surface of the linearized theory is derived and turns out to be manifestly positive definite.

As a result of these investigations, brane induced gravity (BIG) goes without a ghost,
opening an exciting window of opportunity for consistent deformations of gravity
at the largest observable distances.  
\end{abstract}

\maketitle

\section{Introduction}\label{Introduction}
Gravity induced on a surface $\mathcal{M}_4$ of co--dimension $n$ that is equipped 
with a weakly coupled source, embedded in a $d=4+n$--dimensional Minkowski 
space--time $\mathcal{M}_{d}$, is defined as the following effective theory
\begin{eqnarray}
	\label{model}
	\mathcal{S}	
	&=&
	\mathcal{S}_{\rm EH}^{(d)}\left[h\right] + \mathcal{S}_{\rm M}^{(4)}\left[\psi\right]
  	\nonumber \\
	&+& \mathcal{V}_{\rm dyn}^{(4)}[h,\psi] 
	+ \mathcal{V}_{\rm ext}^{(4)}[h]	
	+ \lambda \; \mathcal{S}_{\rm EH}^{(4)}\left[h\right]
\; , 
\end{eqnarray}	
where the first line represents the free graviton action on $\mathcal{M}_{d}$,
and the action describing the free dynamics of all other weakly coupled fields
$\psi$ localized on $\mathcal{M}_4$, respectively. The second line 
collects all effective vertex contributions, i.e.~the minimal coupling of gravitons 
to the dynamically resolved degrees of freedom $\psi$, and the minimal coupling to
external graviton absorbers and emitters. The last term is due to weakly coupled 
fields $\Psi$ that qualify as heavy with respect to some finite cut--off scale
and that have been integrated out in the low energy effective theory (\ref{model}). This  
term is customarily referred to as the induced Einstein--Hilbert term, giving the effective field theory (\ref{model})
its name. In the case of $n=1$ this is the well studied DGP model~\cite{Dvali:2000hr,Dvali:2001ae}. 

It is very important to distinguish between external sources and the induced
Einstein--Hilbert term. The former represent sources that are absolutely 
inert against backreaction, while the latter is, in fact, the dynamical 
fingerprint for the principal presence of $\Psi$ in nature. 

In the literature, the $\mathcal{S}_{\rm EH}^{(4)}\left[h\right]$ contribution
in (\ref{model}) has often been interpreted as an ad hoc kinetic modification rather
than an induced operator. This offered the possibility to probe 
the theory's consistency solely employing external graviton absorbers and
emitters. For $n>1$ it seemed that for phenomenologically interesting choices of the model 
parameters gravitational fluxes originating from the
surface and ending on it violate unitarity. This unitarity violation 
was associated with the gauge invariant scalar carried by propagating
gravitons. From the absorber or emitters viewpoint, this scalar weakened
its own source, while from a dynamical point of view the unitarity violation
manifested itself in a wrong sign for the induced kinetic term, as compared
to the propagation of the transverse and traceless graviton excitations. 
In both cases the unescapable conclusion was that (\ref{model}) exhibits
a ghost~\cite{Dubovsky:2002jm,Hassan:2010ys}. 

However, within the effective field theory framework it is clear that the induced 
Einstein--Hilbert term arises from some weakly coupled fields $\Psi$ at high energies, 
which have been integrated out to obtain the low energy effective field theory (\ref{model}), compare to  \cite{Dvali:2000hr,Dvali:2001ae}. In this way
$\mathcal{S}_{\rm EH}^{(4)}\left[h\right]$ can be understood as a pure source modification. 
In particular, decoupling all $\Psi$ fields implies setting $\lambda$ to zero.
In this case the gravitons
free field theory would be represented by $\mathcal{S}_{\rm EH}^{(d)}$.
So the question arises how a legitimate source could absorb or emit 
a ghost--like excitation?   

Legitimate sources are only required to allow for a Lorentz invariant and gauge invariant 
vertex as well as to respect the usual energy conditions. If unitarity violation would occur, then the source requirements listed above would 
be incomplete, although no further source qualifications are at the core of standard field theory.



Therefore, the physical origin of a ghost--like excitation being emitted or absorbed by
a reasonable source is unclear. To resolve the tension between the physical expectation 
of (\ref{model}) being healthy and the technical analysis of \cite{Dubovsky:2002jm,Hassan:2010ys} 
indicating a ghost in the spectrum is the main motivation for the paper.

The tool that has been used in~\cite{Dubovsky:2002jm,Hassan:2010ys} to diagnose a ghost 
in~(\ref{model}) is the classical brane--to--brane propagator. Its calculation is performed in a manifestly 
covariant framework. Here, the problematic scalar mode contributes a term with a negative 
sign residue. The corresponding quantum propagator is derived by simply applying a Feynman prescription 
to this pole. It causes the vacuum persistence amplitude to grow unbounded, thereby 
threatening the unitarity of~(\ref{model}). 


However, it is found that the $(00)$--Einstein equation, which is a pure constraint, renders the would-be ghost 
mode non-dynamical. Thus, there is no particle interpretation associated with it. This result strongly questions the
reasoning in terms of the classical propagator.

To evaluate the status of (\ref{model}) a full-fledged Hamiltonian analysis was performed incorporating 
all constraints of the system. This allowed to derive the Hamiltonian on the constraint surface which is 
\textit{manifestly positive definite}. This is the main result of this work. It shows that the quantum theory 
is totally consistent, and thus, the problematic scalar mode is not threatening the unitarity of the theory. 
The difficulties in the covariant framework arise because of an inconsistent expression for the quantum 
propagator that is not taking into account the constraints of the theory properly. This problem is closely 
related to the conformal factor problem in GR~\cite{Schleich:1987fm,Mazur:1989by}.


This paper is organized as follows: Section \ref{hamilton_analysis} presents the Hamiltonian 
analysis, with its main result being a positive definite Hamiltonian on the constraint surface. In the first part of Section \ref{Source} the conformal factor problem 
in GR is reviewed and the covariant language is set up. The second part depicts the Lagrangian approach to the model and shows 
that the would-be ghost mode is in fact constrained by the $(00)$--Einstein equation. The proper number of 
sourced propagating degrees of freedom is derived. The conclusion 
is presented in Section~\ref{Summary}, followed by technical details in the appendices.

\section{Hamiltonian Analysis}\label{hamilton_analysis}
In this section the classical Hamiltonian on the constraint surface is calculated and serves as a solid diagnostic 
tool to evaluate the status of \eqref{model}. 
To this end, a complete Dirac constraint analysis is performed for the case
of a two co--dimensional surface $(n=2)$. Since the ghost was formerly derived on a linear level, it suffices 
to study the linear theory of fluctuations $h$ on a Minkowski background.

However, it is convenient to start with the non-linear version of the theory and to decompose the six--dimensional Ricci scalar in its four--dimensional analogue and extrinsic curvature terms. BIG in two co--dimensions is described by the action
 \begin{multline}\label{Theory2}
	\mathcal{S}=\mathcal{S}_{\rm EH}^{(6)}\left[g\right]  + \mathcal{S}_{\rm EH}^{(4)}\left[\omega\right]\\
	=\int \!\!{\rm d}^{6} x \; M_6^4 \sqrt{-g}\, R^{(6)}_{[g]} + \int \!\! {\rm d}^{4} \tilde x\, M_4^2 \sqrt{-\omega}\, R^{(4)}_{[\omega]}
\end{multline}	
where $g$ is the six dimensional bulk metric and $\omega$ the induced metric on the brane. The brane coordinates $\tilde x^{ \alpha}$ are chosen such that they coincide with the bulk coordinates~$ x^{A}$:
$\tilde x^{ \alpha}=\delta^{\alpha}_A x^A$. This static gauge implies $\omega_{\alpha\beta}=\delta_{\alpha}^A\delta_{\beta}^B g_{AB}$.  The index ranges are specified in the table below. $M_6$ is the gravitational scale in the bulk and $M_4$ the induced scale on the brane.  

\subsection{Preparations}
The following canonical analysis uses a multiple ADM-split as starting point. For concreteness, both 
spatial extra--dimensions and the time direction are split in the usual ADM sense:
\begin{align}
	g_{AB} &=\left(\begin{array}{cc} \lambda_{\mu \nu} & \Lambda_{\mu} \\ \Lambda_{\nu} & \Lambda^2 + \Lambda_{\lambda} \Lambda^{\lambda} \end{array}\right) \label{ADM1}
	\, ,
	\intertext{with} 
	\lambda_{\mu \nu}&=\begin{pmatrix} \omega_{\alpha \beta} & \Omega_{\alpha} \\ \Omega_{\beta} & \Omega^2 + \Omega_{\gamma} \Omega^{\gamma} \end{pmatrix}
	\, ,
	\intertext{and}
	\omega_{\alpha \beta}&=\begin{pmatrix} -\Gamma^2 + \Gamma_{i} \Gamma^{i} & \Gamma_{i} \\ \Gamma_{j} & \gamma_{ij} \end{pmatrix}
	\, .
\end{align}

Here, $\gamma_{ij}$ denotes the submetric of the spatial hypersurface orthogonal to the normal vectors 
\begin{align}
	\begin{pmatrix} 
		\hat{n}_6^{A} \end{pmatrix} &= \begin{pmatrix} -\Lambda^{\mu}/\Lambda, & 1/\Lambda 
	\end{pmatrix} 
	\, ,
	\\ 
	\begin{pmatrix} 
		\hat{n}_5^{\mu} 
	\end{pmatrix} 
	&= 
	\begin{pmatrix}
		 -\Omega^{\alpha}/\Omega, & 1/\Omega 
	\end{pmatrix} 
	\, ,\\
	\begin{pmatrix} 
		\hat{n}_4^{\alpha} 
	\end{pmatrix} 
	&= 
	\begin{pmatrix} 
		1/\Gamma, & -\Gamma^i/\Gamma 
	\end{pmatrix} 
	\,,
\end{align}
where $\Lambda$, $\Omega$ and $\Gamma$ denote the three 'Lapse'--functions corresponding to the three ADM--splits, and
$\Lambda^{\mu}$, $\Omega^{\alpha}$ and $\Gamma^i$ are the respective 'Shift'--functions. Indices are raised and lowered 
with the background Minkowski metric~$\eta_{AB}$. The index ranges are as follows:
\\ \\
\begin{tabular}[c]{l|l|c}
	$A$, $B$, $C$, $D$				 	&\,0,\,1,\,2,\,3,\,5,\,6& \\ 
	$\alpha$, $\beta$, $\gamma$, $\delta$\,	&\,0,\,1,\,2,\,3& ADM 4+1\\ 
	$\lambda$, $\mu$, $\nu$, $\rho$\,		&\,0,\,1,\,2,\,3,\,5& ADM 5+1\\ 
	$a$, $b$, $c$, $d$ \;					&\,5,\,6& bulk directions\\ 
	$i$, $j$, $k$, $l$ \;&\,1,\,2,\,3&\; spatial surface directions\\ 
\end{tabular}
\vspace{0.2cm}

The relation between the Ricci scalar of a $d$--di\-men\-sional space--time
and an imbedded space--time with either one temporal or spatial dimension 
reduced is given by, including all total derivative contributions,
\begin{multline}
	\sqrt{-g} R^{(d)} = \sqrt{-g} \Big\{   R^{(d-1)} + (\hat{n}_d \cdot \hat{n}_d) \Big[ (\mathrm{Tr}K_d )^2 - \mathrm{Tr} K_d^2  
	\\
	+ 2 (\nabla \cdot ( (\hat{n}_d \cdot \nabla) \hat{n}_d) - \nabla \cdot (\hat{n}_d (\nabla \cdot \hat{n}_d))) \Big] \Big\} 
	\, . \label{eq_riccisplit}
\end{multline}
Here, the covariant derivative $\nabla$ is compatible with the parent metric $g$ 
characterizing the geometrical structure on the $d$--dimensional space--time. 
In the case of a single ADM--split, the second line in (\ref{eq_riccisplit}) can be dropped 
since it gives rise to a total derivative term in the action.
However, using (\ref{eq_riccisplit}) successively introduces covariant derivatives compatible
with the induced metric on the imbedded space--times, whereas the volume measure in the action
is always given by the parent metric $g$. In this case, the second line in 
(\ref{eq_riccisplit}) needs to be kept. Depending on whether the ADM--split amounts to 
separate either a temporal or a spatial dimension, the product 
$\hat{n}_d \cdot \hat{n}_d$ yields $-1$ or $+1$, respectively. 
$K_d$ denotes the extrinsic curvature tensor in $d$--dimensions. 
In terms of ADM variables,
\begin{align}
	(K_{6})_{\mu \nu} &= \frac{1}{2\Lambda} (\partial_6 \lambda_{\mu \nu} - \nabla_{\mu} \Lambda_{\nu} -  \nabla_{\nu} \Lambda_{\mu}) 
	\, , \\
	(K_{5})_{\alpha \beta} &= \frac{1}{2\Omega} (\partial_5 \omega_{\alpha \beta} - \nabla_{\alpha} \Omega_{\beta} -  \nabla_{\beta} \Omega_{\alpha})
	 \, ,\\
	(K_{4})_{ij} &= \frac{1}{2\Gamma} (\partial_t \gamma_{ij} - \nabla_{i} \Gamma_{j} -  \nabla_{j} \Gamma_{i})
	\, .
\end{align}

\subsection{Calculating the Hamiltonian}
Performing three ADM--splits in succession to separate both spatial extra directions and
the temporal dimension and using (\ref{eq_riccisplit}) in each succession 
yields the following non-linear Lagrangian:
\begin{widetext}
\begin{align}
	\label{eq_lagrangesplit}
	\mathcal L =&\; M_6^4 \sqrt{-g} R^{(6)}_{[g]} + M_4^2 \sqrt{-\omega}\,\delta^{(2)}_y R^{(4)}_{[\omega]}
 	\\ \notag 
 	=&\;\big( M_6^4 \sqrt{-\omega}\, \Omega\, \Lambda + M_4^2 \sqrt{-\omega} \,\delta^{(2)}_y \big)\big\{  R^{(3)}_{[\gamma]} + (K_4)_{ij} (K_4)^{ij} - (K^i_{4 \, i})^2 \big\} 
 	+ M_6^4 \sqrt{-\omega}\, \Omega\, \Lambda \big\{ (K^{\alpha}_{5 \, \alpha})^2-(K_5)_{\alpha \beta} (K_5)^{\alpha \beta}  \big\}\;+& 
  	\\\notag 
  	&\qquad\qquad + M_6^4 \sqrt{-\omega} \, \Omega\, \Lambda \big\{ (K^{\mu}_{6 \, \mu})^2 -(K_6)_{\mu \nu} (K_6)^{\mu \nu}\big\} 
  	- 2 M_6^4 \sqrt{-\lambda}\, \Lambda \bigl\{ \nabla_{\mu} (\hat{n}_5^{\mu} \nabla_{\nu} \hat{n}_5^{\nu} ) -  \nabla_{\nu} (\hat{n}_5^{\mu} \nabla_{\mu} \hat{n}_5^{\nu} ) \big\} \; +&
  	\\ \notag
  	&\quad\qquad\qquad\qquad\qquad\qquad\qquad\qquad\qquad+2 M_6^4 \sqrt{-\omega} \, \Omega\, \Lambda \bigl\{ \nabla_{\alpha} (\hat{n}_4^{\alpha} \nabla_{\beta} \hat{n}_4^{\beta} ) -  \nabla_{\alpha} (\hat{n}_4^		{\beta} \nabla_{\beta} \hat{n}_4^{\alpha} ) \big\} 
	\, .
\end{align}
\end{widetext}
Here, 
$\nabla_{\lambda}$, $\nabla_{\mu}$, $\ldots$ denotes the covariant derivative compatible with the 
induced metric $\lambda_{\mu \nu}$, and $\nabla_{\alpha}$, $\nabla_{\beta}$, $\ldots$ is the covariant 
derivative compatible with $\omega_{\alpha \beta}$.  For ease of notation $\delta^{(2)}_y=\delta^{(2)}(y)$ has been introduced, where $y$ denotes the extra space coordinates $x^5$ and $x^6$, collectively.

Linearizing the space--time geometry around a Min\-kowski background, 
$g_{AB} = \eta_{AB} + h_{AB} $, introduces the following fluctuations for the ADM variables:
\begin{align}\label{LinearizedADM}
	\Gamma &= 1 + \delta \Gamma & \Omega &= 1 + \delta \Omega &   \Lambda &= 1 + \delta \Lambda 
	\notag\\
	&\equiv 1 + n&&\equiv 1 + N&&\equiv 1 + L 
	\notag\\ 
	\notag\\
	\Gamma _i &= \delta \Gamma_i &   \Omega_{\alpha} &= \delta \Omega_{\alpha} &  \Lambda _{\mu} &= \delta \Lambda _{\mu} 
	\notag\\
	&\equiv n_i&&\equiv N_{\alpha}&&\equiv L_{\mu} 
	\notag\\ 
	\notag\\
	\gamma_{ij} &= \delta_{ij} + h_{ij} 
\end{align}

Expanding \eqref{eq_lagrangesplit} up to second order in metric fluctuations, allows to 
derive the canonical momentum fields:
\begin{alignat}{3}
	&\Pi_{n} &&=  \frac{\partial \mathcal L}{\partial \dot n} &&= 0 
	\,,\label{eq_pidn}\\
	&\Pi_{n^i} &&=  \frac{\partial  \mathcal L}{\partial \dot n^i}&&=0 
	\,,\label{eq_pidni}\\
	&\Pi_{N}&&=   \frac{\partial \mathcal L}{\partial \dot N}		
	&& =  M_6^4 \big(- \dot{h}^i_i + 2  \partial_i n^i - 2  \dot L \big) 
	\,,\\
	&\hat{\Pi}_{L} &&= \frac{\partial \mathcal L}{\partial  \dot L} 		
	&&= M_6^4 \big( - \dot{h}^i_i + 2  \partial_i  n^i - 2   \dot N \big)
	\,,\\
	&\Pi_{0} &&= \frac{\partial  \mathcal L}{\partial  \dot N^0}	
	&&=  M_6^4 \big( - \partial_5 h^i_i - 2 \partial_5  L \big) \label{eq_pi0}
	\,,\\
	&\hat{\Pi}_{0} && = \frac{\partial  \mathcal L}{\partial \dot L^0} 
	&&= M_6^4 \big( - \partial_6 h^i_i - 2 \partial_6  N \big) \label{eq_pit0} 
	\,,\\
	&\Pi_i &&= \frac{\partial  \mathcal L}{\partial \dot N^i}		
	&& =  M_6^4 \big( - \partial_5  n_i +  \dot N_i + \partial_i  N^0 \big)
	\,,\\
	&\hat{\Pi}_i &&= \frac{\partial \mathcal L}{\partial \dot L^i} 	
	&&= M_6^4 \big( - \partial_6  n_i +  \dot L_i + \partial_i  L^0 \big)
	\,,\\
	&\hat{\Pi}_{5} &&= \frac{\partial \mathcal L}{\partial  \dot L^5} 	
	&&=  M_6^4 \big(- \partial_6  N_0 +  \dot L_5 - \partial_5  L_0 \big) 
	\,,\\
	&\Pi_{ij} &&= \frac{\partial  \mathcal L}{\partial \dot h^{ij}} 		
	&&=\left(M_6^4 + M_4^2\, \delta^{(2)}_y\right) 
	\Big( \frac{1}{2} \dot{h}_{ij} - \partial_{(i}  n_{j)} \notag 
	\\
	& && && - \frac{1}{2} \delta_{ij} \dot{h}^i_i +\delta_{ij} \partial_k  n^k  \Big)
	- M_6^4  \delta_{ij} \big( \dot N +  \dot L \big) \,.\label{eq_piij} 
\end{alignat}
Since the momentum field \eqref{eq_piij} conjugated to $h_{ij}$
involves a delta function, it can be further decomposed into a regular and an irregular
contribution
$\Pi^{ij} = \Pi^{ij}_{(R)} + \,\delta^{(2)}_y \, \Pi^{ij}_{(I)} $.  
Equations (\ref{eq_pi0}), (\ref{eq_pit0}), (\ref{eq_pidn}) and (\ref{eq_pidni}), constitute 
six primary constraints (in what follows, they will be collectively 
denoted by $\phi^{(1)}_a$, where $a$ runs from $1$ to $6$) as they do not involve 
time derivatives acting on the dynamical degrees of freedom. 
Accordingly, it will not be possible to solve for the 
velocity of the fields $\delta N^0$, $\delta n$, $\delta n^i$ and~$\delta L^0$. 
The velocities of the other fields are given by
\begin{alignat}{2}
	&\dot N_i &&= \frac{1}{M_6^4} \Pi_{ i} + \partial_5 n_i + \partial_i N_0 \label{eq_dotNi} 
	\,,\\
	&\dot L_i &&= \frac{1}{M_6^4} \hat{\Pi}_{i} + \partial_6 n_i + \partial_i  L_0 
	\,,\\
	&\dot L_5 &&= \frac{1}{M_6^4} \hat{\Pi}_{ 5} + \partial_6  N_0 + \partial_5  L_0 
	\,,\\
	&\dot N &&=\!
	\left\{
	\begin{array}{rl}
		\frac{1}{4 M_6^4}\! \left\{ \frac{3}{2} (\Pi_{ N}\! + \hat{\Pi}_{ L}) - \Pi^{\;\;\;\;k}_{(R) \, k} \right\} 
		- \frac{1}{2 M_6^4} \hat\Pi_{ L}\,, \; \vec y \neq 0
		\, ,\\
		\\
		\frac{1}{2 M_4^2}  \Pi^{\;\;\;\;k}_{(I) \, k} - \frac{1}{2 M_6^4} \hat\Pi_{ L}\,,\qquad\qquad\qquad\qquad\; \vec y = 0 
		\,,
	\end{array}
	\right.
	\\
	&\dot L &&=\!
	\left\{
	\begin{array}{rl}
		\frac{1}{4 M_6^4}\! \left\{ \frac{3}{2} (\Pi_{ N}\! + \hat{\Pi}_{ L}) - \Pi^{\;\;\;\;k}_{(R) \, k} \right\} 
		- \frac{1}{2 M_6^4} \Pi_{ N}\,, \; \vec y \neq 0
		\, ,\\
		\\
		\frac{1}{2 M_4^2}  \Pi^{\;\;\;\;k}_{(I) \, k} - \frac{1}{2 M_6^4} \Pi_{ N}\,,\qquad\qquad\qquad\qquad\; \vec y = 0 
		\,,
	\end{array}
	\right.
	\\
	&\dot h_{ij} \!&&=
	\left\{
	\begin{array}{rl}
		\frac{2}{M_6^4} \Pi_{(R) \, ij} - \frac{1}{4 M_6^4} (\Pi_{ N} + \hat{\Pi}_{ L}) \delta_{ij}  \qquad \qquad 
		\\
 		- \frac{1}{ 2M_6^4} \delta_{ij} \Pi_{(R) \, k}^{\;\;\;\;k} + 2 \partial_{(i} n_{j)}\,, \quad \vec y \neq 0
		\, ,\\
		\\
 		\!\!\frac{2}{M_4^2} \Pi_{(I) \, ij}\! - \frac{1}{M_4^2} \delta_{ij} \Pi_{(I) \, k}^{\;\;\;\;k}  + 2 \partial_{(i} n_{j)} \label{eq_dothijy0}\,,\vec y = 0
		\,.
	\end{array}
	\right.
\end{alignat}

Note that the equation determining $\dot L,\, \dot N$ and $\dot h_{ij}$ have been 
decomposed into the two equations, where 
one determines the field value off the surface, 
and the other on the surface localized at $\vec y = 0$. 
This decomposition is a reminiscence of the delta function appearing in (\ref{eq_piij}).
Equations (\ref{eq_dotNi})--(\ref{eq_dothijy0}) allow to derive the Hamiltonian $H$, 
which is a lengthy expression presented explicitly in Appendix \ref{AppH}. 
In order to prove the stability of (\ref{model}), only the Hamiltonian on the constraint surface
is required. 

Canonical consistency demands that the primary constraints are conserved under time evolution. 
This yields the secondary constraints $\phi^{(2)}_a$:
\begin{equation}
	\phi^{(2)}_a = \dot{\phi}^{(1)}_a = \{ H, \phi^{(1)}_a \} 
	\, .
\end{equation}
In greater detail,
\begin{alignat}{2}
	& \phi_1^{(1)} &&= \Pi_0 + M_6^4 \big(  \partial_5 h^i_i + 2 \partial_5 L \big) 
	\,,\\
	& \phi_2^{(1)} &&= \tilde{ \Pi}_0 + M_6^4 \big(  \partial_6 h^i_i + 2 \partial_6 N \big)
	\,,\\
	& \phi_3^{(1)} &&=  \Pi_{n}
	\,,\\
	& \phi_{i+3}^{(1)} &&=  \Pi_{n^i}
	\,,
\end{alignat}
and
\begin{widetext}
\begin{alignat}{2}
	& \phi_1^{(2)} &&= \{ H, \phi_1^{(1)} \} =  \partial_5 \Pi_{ N} + \partial_i \Pi^i + 2 M_6^4 \partial_6 \partial_5 L_0 + \partial_6 \hat{\Pi}_{5} 
	\,,\\
	& \phi_2^{(2)} &&= \{ H,  \phi_2^{(1)} \} =  \partial_6 \hat{\Pi}_{ L} + \partial_i \hat{\Pi}^i 
	+ 2 M_6^4 \partial_6 \partial_5 N_0 + \partial_5 \hat{\Pi}_{5} 
	\,,\\ 
	& \phi_3^{(2)} &&= \{ H,  \phi_3^{(1)}\} =  M_6^4 \big\{ ( \partial_5^2  + \partial_6^2)  h^i_i - 2 \partial_6 \partial_i L^i - 2 \partial_5 \partial_i N^i - 2 \partial_5 \partial_6 L_5 \\*
	& &&\qquad\qquad\qquad\qquad\qquad\qquad\qquad+ 2 \partial_i \partial^i ( N +  L)  + 2 \partial_6^2 N+ 2 \partial_5^2 L  \big\}- (M_6^4+M_4^2 \delta^{(2)}_y) \delta^{1}\!R^{(3)} 
	\label{HamiltonianConstraint}
	\, ,\\ 
	& \phi_{i+3}^{(2)} &&= \{ H,  \phi_{i+3}^{(1)} \} = - 2 \partial_j \Pi^{ij} - 2 M_6^4 ( \partial_6 \partial^i L_0 + \partial_5 \partial^i N_0 ) - \partial_5 \Pi^i - \partial_6 \hat{\Pi}^i 
	\, .
\end{alignat} 
\end{widetext}
Here, $\delta^{1}\!R^{(3)}=\partial_i\partial_j h^{ij}-\partial_i\partial^i h_k^k$ is the first 
order variation of the Ricci-scalar for the surface's spatial dimensions. 
$\phi_3^{(2)}$ denotes the generalization of the so--called Hamiltonian constraint 
in general relativity, which can be seen by sending $M_6^4$ to zero. It is equivalent to 
the constraint given by the (00)--Einstein equation, compare to Section \ref{Source}. 
These secondary constraints are all conserved under time evolution, 
i.e.~they commute with the Hamiltonian under the Poisson bracket
\begin{equation} \label{Tertiary}
	\dot \phi_a^{(2)} =  \{ H,  \phi_a^{(2)}\} \simeq 0 \, .
\end{equation}
These relations have been checked explicitly. Note that the last relation $(\simeq)$ is a weak equality, which means that the right hand side of (\ref{Tertiary}) is a linear combination of the constraints $\phi^{(p)}_a$ of the system ($p\in\{1,2\}\;\& \; a \in \{1,2,3,4,5,6\}$).
According to \eqref{Tertiary}, the system does not possess any tertiary constraints.
Thus, the constraint content is given by 
the set of $12$ primary and secondary constraints $\phi^{(p)}_a$.
Moreover, it can be shown that the constraint system is completely first class:
\begin{align}
	\forall \;p,p' , a, a': \;\; \{ \phi^{(p)}_a, \phi^{(p')}_{a'}\} \simeq 0
	\, .
\end{align}
As a consequence, every constraint generates a gauge transformation on any
quantity $\Theta$, that is built up out of the dynamical field variables, as follows:
\begin{equation}
	\delta \Theta = \xi \{ \Theta, \phi^{(p)}_a \} \, ,
\end{equation}
with a space--time dependent gauge function $\xi$. 
In this way, the 12 first class constraints allow to reduce the number
of independent dynamical degrees of freedom by~24. 
This freedom allows to implement the gauges:
\begin{alignat}{2}
	& \phi_1^{(1)}&&:\;  N_0 = 0 
	\; ,\label{gauge1}\\
	& \phi_2^{(1)}&&:\;  L_0 = 0
	 \, ,\\
	& \phi_3^{(1)}&&:\;  n = 0 
	\, ,\\
	& \phi_{3+i}^{(1)}&&:\;  n^i = 0
	 \, ,\\
	& \phi_2^{(2)} &&:\;  L_5 = 0 
	\, ,\label{gauge4}\\
	& \phi_3^{(2)} &&:\; \Pi^i_i = 0 
	\, ,\\
	& \phi_{3+i}^{(2)} &&:\; \partial_j h^{ij} = 0
 	\,.
	\label{gauge2}
\end{alignat} 
The gauge freedom given by $\phi_1^{(2)}$ will be used later.

Because all the sources in the BIG setup are four-dimensional and localized on the higher co--dimensional surface, 
graviton absorption and emission processes
are spatially isotropic in the directions normal to it. This means that the 
graviton field $h$ respects a $SO(2)$--symmetry with respect to the two extra dimensions.
Moreover, the derivation of the would-be ghost in former works is solid under the assumption of such a symmetry.
Thus, in order to show the absence of the ghost, it is justified to make use of this symmetry.
It is most easily implemented using polar coordinates 
$(r,\varphi)$, where $x^5=r \cos\varphi$ and  $x^6=r\sin\varphi$. Then the symmetry 
demands the extra space components of the graviton field not to depend on
~$\varphi$. Additionally, the $h_{\varphi r}$ components have to vanish. Transforming 
back to Cartesian coordinates yields
\begin{align}
	N&=\cos^2{\!\varphi}\, h_{rr}+\frac{\sin^2{\!\varphi}}{r^2}\, h_{\varphi\varphi}
	\,,\\
	L&=\sin^2{\!\varphi}\, h_{rr}+\frac{\cos^2{\!\varphi}}{r^2}\, h_{\varphi\varphi}
	\,,\\
	L_5&=\cos{\varphi}\sin{\varphi}\,h_{rr}-\frac{\cos{\varphi}\sin{\varphi}}{r^2}\,h_{\varphi\varphi}
	\,.
\end{align}
The gauge choice \eqref{gauge4} then implies $r^2\,h_{rr}=h_{\varphi\varphi}$, which in turn 
demands 
\begin{align}
	N &= L
	\, ,\label{NeqL}
\end{align}
where $N(r,x)$ only depends on $r$ and $x$. The symmetry implies as well $h_{\varphi j}=0$.
Using the same reasoning as before, one finds
\begin{align}\label{sym2}
	 N_i&= \tilde{N}_i\cos\varphi &\rm{and} && L_i= \tilde{N}_i \sin\varphi \,,
\end{align}
where $\tilde{N}_i(r,x)$ is a function of $r$ and $x$. Similarly, for the $\Pi$-sector,  
\begin{align}
	&\Pi_{N} = \hat{\Pi}_{L}
	\,,\label{sym3}	\\
	&\Pi_i= \tilde{\Pi}_i \cos\varphi &\rm{and} &&\hat{\Pi}_i= \tilde{\Pi}_i\sin\varphi 
	\,,\label{sym4}	\\
	&\hat{\Pi}_5=0
	\,.\label{sym5}
\end{align}
Here, $\Pi_{N}$ and $\tilde{\Pi}_i$ are $\varphi$--independent. 
It should be noticed that these relations are gauge dependent.  
The remaining gauge freedom corresponding to $\phi_1^{(2)}$ can be used to implement
\begin{align} \label{gauge3}
	\partial_r \partial_i \tilde{N}^i=(\Delta_3+\Delta_2) N\,.
\end{align}
In this gauge, the  Hamiltonian constraint $\phi_3^{(2)}$ simplifies to
\begin{align}\label{constraintH2}
	\tilde{\Delta}h_i^i =0\,,
\end{align}
with the generalized Laplace operator defined as
\begin{align}
	\tilde{\Delta}\equiv\left[\Delta_2+\Delta_3 + \frac{M_4^2}{M_6^4}\delta^{(2)}_y\;\Delta_3\right] \,.
\end{align}
The only bounded solution to the constraint equation~(\ref{constraintH2}) is $h_i^i=0$ . 

Using all gauge conditions, \eqref{gauge1}--\eqref{gauge2}, as well as \eqref{NeqL}, \eqref{gauge3} and 
\eqref{constraintH2}, the Hamiltonian on the constraint surface is given by
\begin{widetext}
\begin{multline}\label{FinalH}
	\mathcal{H}=\frac{1}{M_6^4} \Pi^{(T)}_{(R)ij}\Pi^{(T)ij}_{(R)}+\frac{1}{M_4^2} \,\delta^{(2)}_y\, \Pi^{(T)}_{(I)ij}\Pi^{(T)ij}_{(I)} + \frac{1}{4 M_6^4} \Pi^2_{ N}+\frac{1}{2 M_6^4} \tilde{\Pi}_i\tilde{\Pi}^i + 		\frac{1}{4} M_6^4 \tilde{F}_{ij} \tilde{F}^{ij} 
	\\*
	+ \frac{1}{4} M_6^4 \partial_a h_{ij}^{(tt)}  \partial^a h^{(tt) ij} + \frac{1}{4} \left(M_6^4 + M_4^2 \,\delta^{(2)}_y\right) \partial_k h_{ij}^{(tt)}  \partial^k h^{(tt) ij} +2 M_6^4 \partial_a N \partial^a  N 
	\, .
\end{multline}
\end{widetext}
Here, 
$\tilde{F}_{ij}=\partial_i \tilde{N}_j-\partial_j \tilde{N}_i$, 
$\Pi_{ij}^{(T)}$ denotes the traceless part of the momentum field $\Pi_{ij}$ 
and $h_{ij}^{(tt)}$ is the transverse and traceless part of $h_{ij}$. 

Evidently, $\mathcal{H}$ consists only of positive squares, which 
implies that the Hamiltonian $H$ is positive definite,
which in turn is a sufficient condition for a ghost--free theory. Note that 
a real ghost degree of freedom, which originates from a negative sign kinetic 
operator, would inevitably destroy the positive definiteness of the 
classical Hamiltonian. It should be stressed that this result does
not depend on the fact of having a perfectly localized brane. $\delta_y^{(2)}$ 
should rather be thought of some finite width profile function.

\subsection{Counting Degrees of Freedom}
It is very instructive to count the degrees of freedom in
the effective theory (\ref{model}) for a two co--dimensional
source. The source--free theory is gravity in six dimensions,
which suggests $[h]\le 9$. Note that it could be less than nine due 
to the fact that the source is four--dimensional and brane--localized. This introduces an additional $SO(2)$--symmetry
by which the number of independent degrees of freedom gets reduced. Moreover,
some of the potential degrees of freedom could turn out not to be sourced. 

 Dirac's constraint analysis 
allows to do a solid counting. Because of the index symmetries,
$[h]\le 21$, which is doubled in phase space due 
to the conjugated momentum fields. The dynamical system 
possesses 12 first class constraints, each generating
a gauge transformation. This allows to remove 24 gauge 
redundancies, leaving $[h]\le 9$, which is again doubled in
phase space. This is already confirming the naive assumption
relying on the effective field theory argumentation.

The list containing the remaining conjugated 
pairs is of course a gauge dependent statement. 
For instance, in what follows, the gauge freedom represented by 
$\phi_{3}^{(2)}$ is used to render $\Pi_{ij}$ traceless,
whereas the longitudinal part of $\Pi_{ij}$
is fixed employing the constraint  $\phi_{3+i}^{(2)}$.
As a consequence, the conjugated momentum field in this gauge
becomes $\Pi_{ij}^{(tt)}$, which is the transverse and traceless part 
of $\Pi_{ij}$, and similarly for the other fields. A possible list after 
gauge fixing is:
\\\\
\begin{tabular}[c]{c|c|c}
	Conjugate pairs \; & Degrees of freedom\; & Constraint 	
	\\ \hline
	$(h^{(tt)}_{ij},\Pi^{(tt)}_{ij})$ \; 	& \; 2 & $\phi_{3+i}^{(2)}$, $\phi_{3}^{(2)}$ 	
	\\ 
	$(N,\Pi_N)$ \;&\; 1 &	\;
	\\ 
	$(L,\hat{\Pi}_L)$ \;&\; 1		&
	\\ 
	$(N_i^{(t)},\Pi_i^{(t)})$ \;&\; 2	&	$\phi_{1}^{(2)}$ 
	\\ 
	$(L_i,\hat{\Pi}_i)$ \;&\; 3 		&	
\end{tabular}
\vspace{0.2cm}

The background isometry $SO(2)$ in the space of directions transverse to the surface
reduces this list further. Taking \eqref{sym2}--\eqref{sym5} into account, it becomes:

\begin{tabular}[c]{c|c|c}
	Conjugate pairs \; & Degrees of freedom\; & Constraint 	
	\\ \hline
	$(h^{(tt)}_{ij},\Pi^{(tt)}_{ij})$ \; & \; 2 & $\phi_{3+i}^{(2)}$, $\phi_{3}^{(2)}$ 
	\\ 
	$(N,\Pi_N)$ \;&\; 1 &			
	\\ 
	$(\tilde{N}_i^{(t)},\tilde{\Pi}_i^{(t)})$ \;&\; 2	&	$\phi_{1}^{(2)}$ 
\end{tabular}
\vspace{0.2cm}

The canonical counting gives five dynamical degrees of freedom.
This number agrees with the degrees of freedom in the Dvali--Gabadadze--Porrati model,
which corresponds to (\ref{model}) for the special case of one co--dimension. 
There it has been shown that the graviton is a continuous superposition of
massive spin--2 excitations each propagating five helicity components. 
However, as it will become more clear in Section \ref{Source}, only $h_{ij}^{(tt)}$ can be sourced by a localized four-dimensional source. 
Therefore, the number of sourced degrees of freedom is given by $[h^{(tt)}]=2$ coinciding 
with the results found in the covariant analysis below.

\section{Semi--covariant analysis} \label{Source}
In \cite{Dubovsky:2002jm,Hassan:2010ys} a covariant language was used to derive the ghost in the BIG model. The appearance of this unitarity violating mode clearly contradicts the result of Section~\ref{hamilton_analysis}. In order to make contact to these works, the covariant approach is studied in detail here. An explanation is offered why the covariant treatment does not allow to reliably address the unitarity issue. The main argument boils down to the statement that the scalar which was always regarded as a threat to unitarity is not dynamical. Former works did not take into account this constraint nature of the scalar mode properly, since both,~\cite{Dubovsky:2002jm} and \cite{Hassan:2010ys}, indicate that a ghost-like scalar degree of freedom can be found amongst the physical particle content of the theory.  
Furthermore, in this chapter a more physical viewpoint is established in which the BIG term plays the role of a localized source modification. The source argument
provides a physical indication that the theory should be healthy.

\subsection{Conformal factor problem in GR} \label{cfpGR}
The seeming unitarity violation in BIG is known to be mediated by the conformal mode of the graviton. 
In GR there is a very similar problem, sometimes called the conformal factor problem. Here the conformal 
mode of the graviton is threatening the unitarity of the corresponding quantum theory, too. In GR the appearance 
of this problem is strongly tied to the covariant description of the system and absent in the canonical formulation. 
One important aim of this work is to show that both problems, the one in GR and the one in BIG, are closely related. 
As a warming up exercise and in order to set up the covariant language, the conformal factor problem in standard 
GR in a weakly coupling regime is studied first.

Let $(\mathcal{M}_{_4},\eta)$ be a four--dimensional space--time, equipped with a geometrical structure provided by the Minkowski 
metric $\eta$. The action reads
\begin{align}
	\mathcal{S}=\mathcal{S}_{\rm EH}^{(4)}\left[h\right]  + \mathcal{V}^{(4)}[h]\;,
\end{align}	
where $\mathcal{S}_{\rm EH}^{(4)}\left[h\right]$ is the perturbed Einstein-Hilbert action on a Minkowski background 
up to second orders in $h$. A source can absorb and emit gravitons $h$, respectively, 
according to the minimal coupling vertex 
\begin{equation}
	\mathcal{V}^{(4)}
	= 
	\int_{\mathcal{M}_{_4}}\!\!\!\!\! {\rm d}^{4} x
	\; h_{\alpha\beta} \; t^{\alpha\beta}
	\; .
\end{equation}	
In order to further study the dynamics of this model, the graviton $h$ is decomposed into its gauge invariant and 
gauge variant contributions
\begin{align}\label{4DDecomp}
	h_{\alpha\beta}
	=D^{(\rm tt)}_{\;\;\; \alpha\beta} + \partial_{(\alpha} V^{_{(\perp)}}_{\;\;\; \beta)} +
	P^{_{(\parallel)}}_{\;\; \alpha\beta} \; B + \eta_{\alpha\beta} \; S
	\;,
\end{align}
where $D^{(tt)}$ is the transverse and traceless tensor part of the graviton and $V^{_{(\perp)}}$ denotes it 
transverse vector part. $B$ and $S$ are the gauge variant and gauge invariant scalar parts, respectively. 
The field $S$ is the aforementioned conformal mode. The corresponding projectors are specified in the 
Appendix \ref{deconstruction}. The Einstein equations then are
\begin{align} \label{EinsteinGR}
	\Box_4 D^{(\rm tt)}_{\;\;\;\alpha\beta}-2\eta_{\alpha\beta}\Box_4 S + 2 \partial_{\alpha}\partial_{\beta}S
	=-2\kappa_0 \,t_{\alpha\beta}
	\,,
\end{align}
where the pure gauge modes $V^{_{(\perp)}}$ and $B$ dropped out and $\Box_4 = \eta^{\alpha\beta}\partial_\alpha \partial_\beta$. 
$\kappa_0$ denotes the gravitational strength with which gravitons couple to a conserved source. 
The energy momentum tensor $t$ is decomposed into its transverse--traceless part~$t^{(\rm tt)}$ and its trace~$t_\alpha^\alpha$,
\begin{align}\label{DecomT}
	t_{\alpha\beta}=
	t^{(\rm tt)}_{\;\;\;\alpha\beta}-\frac{1}{3}P^{_{(\parallel)}}_{\;\; \alpha\beta}\,t_{\gamma}^{\gamma}+\frac{1}{3}\eta_{\alpha\beta}\,t_{\gamma}^{\gamma}
	\,.
\end{align}
Taking the trace and applying the transverse-traceless projector, respectively, allows to decompose the Einstein 
equations \eqref{EinsteinGR}
\begin{align}
	+ \; \Box_4 S&=
	\frac{\kappa_0}{3}\;t_\alpha^{\; \alpha} \label{GRS}
	\;,
	\\
	-\;\Box_4 D^{(\rm tt)}_{\;\;\;\;\alpha\beta}&=
	2\kappa_0\; t^{(\rm tt)}_{\;\;\;\;\alpha\beta}\label{GRD}
	\;. 
\end{align}
These equations suggest that the scalar mode $S$ is a ghost as it comes with a different sign for its kinetic 
term compared to $D^{(\rm tt)}$. 

A slightly different way to phrase the problem consists in considering the lorentzian functional integral of the free theory
\begin{align}\label{PathIntGR}
	\int\mathcal{D}\left[h\right]e^{i\mathcal{S}_{\rm EH}^{(4)}\left[h\right]}
	\;.
\end{align}
Expressed in terms of the decomposition \eqref{4DDecomp} the real time lorentzian action is
\begin{align}\label{actionGR}
	\kappa_0\, \mathcal{S}_{\rm EH}^{(4)}\left[h\right]=
	\frac{1}{4}\int_{\mathcal{M}_{_4}}\!\!\!\!\! {\rm d}^{4} x \!
	\left[
		-\left(\partial_\gamma D^{(tt)}_{\alpha\beta}\right)^2 + 6 \left(\partial_{\gamma} S\right)^2  
	\right]
	\,.
\end{align}
 
Performing the euclidian continuation $t\rightarrow-i\tau$ yields an expression for the corresponding euclidian path integral. 
As the kinetic term of the conformal mode $S$ in~\eqref{actionGR} comes with the wrong sign, this integral is divergent 
and hence ill defined.  This so-called conformal factor problem is for example discussed in~\cite{Schleich:1987fm,Mazur:1989by} in the case of GR. 

Of course, it is known that linearized GR on a Minkowski background is a healthy quantum theory. This can be proven by 
performing a Dirac constraint analysis showing that the Hamiltonian is a positive definite quantity, which is a sufficient 
condition for a theory to respect unitarity. This is exactly what is predicted by the positive energy theorem for asymptotically flat space-times~\cite{Witten:1981mf,Schon:1979rg}. So the question naturally arises, why the above analysis is suggesting a 
different result. An answer was given in~\cite{Schleich:1987fm,Mazur:1989by}: The conformal mode $S$ is no independent degree of freedom. It is constraint by the physical degrees of freedom which are contained in $D^{(\rm tt)}$ and the matter sector. Therefore, the $S$ mode cannot be a ghost as there is no state in the Hilbert space associated with it. In the case of the path integral the summation is only allowed to include the true physical degrees of freedom.

In order to reveal the constraint, it is necessary to abandon the manifestly covariant description of the system and depict the $(00)$--Einstein 
equation. Using the transversality of $D^{(\rm tt)}$ and its traceless property, the $(00)$--component of \eqref{EinsteinGR} is
\begin{align}\label{constraintGRS}
	\Delta_3D_{\;\;\;i}^{(\mathrm{tt})i}-\partial^i\partial^jD_{\;\;\;ij}^{(\rm tt)}+2\Delta_3S
	=-2\kappa_0 \,t_{00}
	\,,
 \end{align}
 where $\Delta_3 = \delta^{ij}\partial_i \partial_j$, with $i, j$ running over the spatial directions. As there are no time 
 derivatives occurring in this equation, it is a constraint. Note that $D^{(\rm tt)}$ contains two physical degrees of freedom that are encapsulated 
 in its $(ij)$--components. The transversality of $D^{(\rm tt)}$  is constraining the $(0\beta)$--components. The 6 independent $(ij)$--components 
 are further reduced by the traceless property and the $(0j)$--Einstein equations that are constraints on~$D^{(\rm tt)}$ as well. The dynamics of these 
 remaining two degrees of freedom is fully captured by equation~\eqref{GRD}. Once the dynamical equation for $D^{(\rm tt)}$ is solved, the conformal 
 mode is totally fixed by equation~\eqref{constraintGRS}. 

One might ask whether this is indicating an inconsistency between the dynamical \eqref{GRS} and the constraint \eqref{GRD} equation already on a classical level. However, 
it can be explicitly shown that the solution $S$ of the constraint solves the would-be dynamical equation: Substituting the solution of $D^{(\rm tt)}$ 
in \eqref{constraintGRS} and using the decomposition of the energy momentum tensor \eqref{DecomT} yields 
\begin{align}\label{solutionSGR}
	\Delta_3S=\frac{\kappa_0}{3} \Delta_3\frac{1}{\Box_4}t_{\alpha}^{\alpha}
	\,.
\end{align}
By demanding that $S$ should fulfill appropriate fall-off conditions at spatial infinity, the $\Delta_3$--operator may be simply dropped. The solution for $S$ then obviously solves the would-be dynamical equation \eqref{GRS}. It is clear that the non-local operator~$1/\Box_4$ arises, because a solution to the equations of motion~\eqref{GRD} was inserted in the constraint~\eqref{constraintGRS}. 
The purpose of this GR exercise was to show that a solid analysis of the unitarity issue necessitates to first extract the true dynamical content of the theory. This however is not possible in a completely covariant description and requires to study the $(00)$--component of the Einstein equation. 
It should be noted that the canonical hamiltonian description of the free theory is well defined and does not bear these difficulties. Once the Hamiltonian on the constraint surface is bounded from below, the theory is in accordance with unitarity. 

In Section \ref{LocalizedS} the same semi-covariant arguments are presented for BIG in order to make contact to former works that were using the same language~\cite{Dubovsky:2002jm,Hassan:2010ys}. In a first step ordinary GR in higher dimensions with a localized four--dimensional source is investigated. This allows to establish the higher dimensional framework and highlight the source arguments. In the next step this scenario is generalized to BIG. Equivalently to the GR case the worrisome $S$ mode is not dynamical and therefore does not constitute a ghost. 

\subsection{Gravity of a localized source} \label{LocalizedS}

Let $(\mathcal{M}_{_d},\eta)$ be a $(d=4+n)$--dimensional space--time,
equipped with a geometrical structure provided by the Minkowski 
metric $\eta$. Embedded in this space--time is a source of co--dimension $n$,
localized on a background geometry $(\mathcal{M}_{_4},\eta)$. 
The presence of the source makes it natural to consider the space--time
isomorphism $(\mathcal{M}_{_d},\eta)\cong (\mathcal{M}_{_4},\eta)\times (\mathbb{R}^n,\delta)$,
where $\mathbb{R}^n$ denotes an $n$--dimensional vector space with Euclidean geometry $\delta$. 
A coordinate system covering $(\mathcal{M}_{_d},\eta)$ can be described as follows:
$\eta=\eta_{AB}\,{\rm d}X^A\otimes{\rm d}X^B=
\eta_{\alpha\beta}\;{\rm d}x^{\alpha}\otimes {\rm d}x^{\beta}+
\delta_{ab}\;{\rm d}y^{a}\otimes {\rm d}y^{b}$, with obvious index ranges. 
Then, the localized source is given by
\begin{equation}
	\label{locsour}
	T_{AB}(X)
	= L_{AB}^{\; \; \; \; \; \alpha\beta}(y_0,y) \; t_{\alpha\beta}(x,y)
	\; ,
\end{equation}
where $L(y,y_0)$ denotes a localizer concentrating the sources energy--momentum
around a fixed position $y_0\in\mathbb{R}^n$ in the directions transverse
to $(\mathcal{M}_{_4},\eta)$. In gauge theories the localizer density must allow
for a conserved source.     

The action reads
\begin{align}\label{ActionGRSource}
	\mathcal{S}=\mathcal{S}_{\rm EH}^{(d)}\left[h\right]  + \mathcal{V}[h]\;,
\end{align}	
where $\mathcal{S}_{\rm EH}^{(4)}\left[h\right]$ is the second order, perturbed Einstein-Hilbert 
action on a Minkowski background in $d$ dimensions.

The source can absorb and emit gravitons $h$, respectively, 
from and in all space--time directions according to the 
minimal coupling vertex 
\begin{equation}
	\mathcal{V}
	= 
	\int_{\mathcal{M}_{_4}}\!\!\!\!\! {\rm d}^{4} x \; \int_{\mathbb{R}^n}\!\!\! {\rm d}^{n} y
	\; h_{AB} \; T^{AB}
	\; .
\end{equation}	
If the localizer is ideal, i.e.~a distribution describing a sharp source extension
in the transverse directions, the vertex density becomes effectively four--dimensional, and 
Lorentz--invariance requires only to integrate over $\mathcal{M}_{_4}$ at~$y_0$.
Invariance under gauge transformations requires a conserved localized source,
which in turn implies 
\begin{equation}
	\label{vertex}
	\mathcal{V}
	=
	\int_{\mathcal{M}_{_4}}\!\!\!\!{\rm d}^{4} x\left(
	\; D^{\rm (tt)}_{\; \; \; \; \alpha\beta} \; t^{{\rm (tt)}\; \alpha\beta} +
	S t_\gamma^{\; \gamma}\right)(x,y_0) 
	\; ,
\end{equation}	
where the decomposition \eqref{4DDecomp} has been used for the $(\alpha\beta)$--components 
of the graviton. It follows that the 
source specifications allow for no more than six sourced degrees 
of freedom, distributed as follows: $\left[D^{(tt)}\right]=5$
and $\left[S\right]=1$. Although this counting applies to gauge invariant
objects, the theory may contain further constraints that reduce the number
of dynamical degrees of freedom. Exactly as in the case of four--dimensional GR,
 discussed before. Therefore, at this stage, the number of 
sourced and propagating degrees of freedom is~$\left[h_{\alpha\beta}\right]\le 6$.

The graviton flux is non--vanishing away from the source anchored at $y_0$. 
In order to extract the dynamical content of the theory, this necessitates
to deconstruct the entire graviton $h$ in its gauge invariant and 
gauge variant contributions. In a particular
coordinate system, outlined in the Appendix, the gauge fixed deconstruction is
given by
\begin{eqnarray}
	\label{deco}
	h_{\alpha\beta}
	&=&
	D^{(\rm tt)}_{\;\;\;\; \alpha\beta} + 
	P^{_{(\parallel)}}_{\;\; \alpha\beta} \; B + \eta_{\alpha\beta} \; S
	\; , \nonumber \\
	h_{ab}
	&=&
	d^{(\rm tt)}_{\;\;\;\; ab} + \delta_{ab} \; s
	\; , \nonumber \\
	h_{\alpha b}
	&=&
	G^{(\rm v,v)}_{\; \; \; \; \alpha b} + \partial_b G^{(\rm v,s)}_{\;\;\;\; \alpha}
	+ \partial_\alpha F^{(\rm s,v)}_{\;\;\;\; b}
	\; .
\end{eqnarray}
All tensor qualifications are with respect to the isometries underlying
the space--time $(\mathcal{M}_{_4},\eta)\times(\mathbb{R}^n,\delta)$.
Thus, $D^{(\rm tt)}$ and $B, S$ transform under Lorentz transformations $SO(1,3)$ as 
a transverse and traceless second rank tensor 
and two scalars, respectively, while $d^{(\rm tt)}$ and
$s$ transform under the roation group $SO(n)$, operating on the transverse directions,
as a transverse and traceless second rank tensor and a scalar, respectively. 
The mixed sector involves 
$G^{(\rm v,v)}$, $G^{(\rm v,s)}$ and $F^{(\rm s,v)}$, which transform 
under the direct product $SO(1,3)\times SO(n)$
as (transverse vector, transverse vector), (transverse vector, scalar) and 
(scalar, transverse vector) quantities, respectively.
More details can be found in the Appendix. 

The a priori gauge invariant contributions are $D^{(\rm tt)}$, $S$, $d^{(\rm tt)}$, $s$, $G^{(\rm v,v)}$,
while $B$, $G^{(\rm v,s)}$, $F^{(\rm s,v)}$ resemble gauge fixed quantities. 
Before fixing the gauge, for an extended source and $n>1$, 
$\left[h_{AB}\right]\le (5+n)(4+n)/2$, distributed as follows:
$\left[h_{\alpha \beta}\right]\le 10$, $\left[h_{ab}\right]\le (n+1)n/2$ and 
$\left[h_{\alpha b}\right] \le 4n$. After eliminating the gauge redundancies,
$h_{AB}$ carries no more than $(4+n)(3+n)/2$ propagating degrees of freedom. 
In detail, $h_{\alpha \beta}$ carries no more than seven degrees of freedom,
where $\left[D^{(\rm tt)}\right]\le 5$, $\left[B\right]=\left[S\right]\le 1$,
$h_{ab}$ carries no more than $n(n-1)/2$ propagating degrees of freedom,
where $\left[d^{(\rm tt)}\right]\le n(n-1)/2-1$, $\left[s\right]\le 1$, and,
finally, $h_{\alpha b}$ carries no more than $4n-1$ dynamical degrees of freedom,
where $\left[G^{(\rm v,v)}\right]\le 3(n-1)$, $\left[G^{(\rm v,s)}\right]\le 3 $
and $\left[F^{(\rm s,v)}\right]\le n-1$. This number gets further reduced to
$\left[h_{AB}\right]\le (4+n)(1+n)/2$ by the existence of $d$ constraints encapsulated 
in the $(A\,0)$--Einstein equations. These remaining degrees of freedom are all truely dynamical 
but not necessarily sourced. 

Given the source specifications outlined above, the observable quanta requiring
a unitary evolution correspond to a subset of the sourced fields $D^{(\rm tt)}$ and $S$. 
For these, the dynamical equations are given by
\begin{eqnarray}
	\label{eomss1}
	&&\Box S 
	= 
	-\frac{2}{3} \; \kappa_n \; \frac{n-1}{n+2} \; \widehat{t}_\gamma^{\; \gamma}   
	\; , \\ \nonumber \\
	\label{eomss2}
	&&\Box D^{(\rm tt)}_{\;\;\;\;\alpha\beta}
	=
	-2 \; \kappa_n \; \widehat{t}^{(\rm tt)}_{\;\;\;\;\alpha\beta} 
	\; ,
\end{eqnarray}
where 
$\Box=\Box_4 + \Delta_n$, $\Box_4 = \eta^{\alpha\beta}\partial_\alpha \partial_\beta$,
$\Delta_n = \delta^{ab}\partial_a \partial_b$, and $\kappa_n$ denotes the gravitational 
strength with which gravitons couple to a conserved source in $d=4+n$. It is related to the 
scales used in Section \ref{hamilton_analysis} by $M_{4+n}=\kappa_{n}^{-1/(2+n)}$.
For ease of notation, $\widehat{t}\equiv t \; \delta^{(n)}(y-y_0)$ has been introduced.
All other fields that are left after eliminating the gauge redundancies,
$B, d^{(tt)}, s, G^{(v,v)}, G^{(v,s)}$ and $F^{(s,v)}$ are decoupled from the source.
In the general case, a source extended in the transverse directions would couple to
$d^{(tt)}, s, G^{(v,v)}$ with equal strength $\kappa_n$, while $B, G^{(v,s)}$ and
$F^{(s,v)}$ are always decoupled by gauge invariance. For example, by applying the 
projector for the transverse vector ($SO(1,3)$) on the ($\alpha\beta$)--Einstein equations,
it follows
\begin{align}
\Delta_n G^{(v,s)}_{\alpha}=0\;.
\end{align}

In order to extract the true dynamical content of \eqref{ActionGRSource}, (\ref{eomss1}, \ref{eomss2}) have to be supplemented with the 
$(00)$--Einstein equation, which imposes a constraint on the dynamics 
of $D^{(\rm tt)}$ and $S$. This equation is
\begin{multline}\label{00Ein}
	\left[\partial^i\partial^j\!-\delta^{ij}(\Delta_3\!+\!\Delta_n)\right] D_{\;\;\;ij}^{(\rm tt)}\!-\Delta_nP^{_{(\parallel)}i}_{\;\;i} B \;  
	\\
	\qquad-\left[2\Delta_3+3\Delta_n\right]S-\left[n\,\Delta_3+(n-1)\Delta_n \right]s
	=2\kappa_n \,\widehat t_{00}
	\;,
\end{multline}
where the following definitions apply: $\Delta_3 = \delta^{ij}\partial_i \partial_j$, with $i, j$ running 
over the spatial directions along the brane, and $P^{_{(\parallel)}i}_{\;\;i}$ is the trace over the spatial 
components of the longitudinal projector on $(\mathcal{M}_4,\eta)$. Note that there are no time 
derivatives occurring in this equation. A more rigorous analysis of (\ref{00Ein}) requires to study the equations for 
$s$ and $B$. An appropriate linear
combination of the trace and longitudinal--longitudinal part of the $(\alpha\beta)$--Einstein equations
yields
\begin{align}\label{S_s}
	\Delta_nS=-\frac{(n-1)}{3} \Delta_ns
	\;.
\end{align}
It follows that $S=-(n-1)s/3 + \Phi$, where $\Phi$ is a solution of the Laplace equation, $\Delta_n \Phi=0$, and has to be taken to be zero as the fields have to respect appropriate fall-off conditions. By inserting this in the $(ab)$--Einstein equations describing the longitudinal--longitudinal dynamics one finds
\begin{equation}\label{B_S}
	\Delta_n B
	=
	-\frac{n+2}{n-1} \Delta_n S
	\;.
\end{equation}
Equation \eqref{00Ein} becomes
\begin{multline}\label{ConstraintS}
	\left[\partial^i\partial^j\!-\delta^{ij}(\Delta_3\!+\!\Delta_n)\right] D_{\;\;\;ij}^{(\rm tt)}\! +\!\frac{n+\!2}{n-\!1}\left[\Delta_nP^{_{(\parallel)}i}_{\;\;i}\!\! +\!\Delta_3\right]\!S\\
	=2\,\kappa_n \,\widehat t_{00}
	\;.
\end{multline}
 
Equation \eqref{ConstraintS} does not contain any time derivatives and therefore 
is a constraint on $S$, that is sometimes referred to as the Hamiltonian constraint.  
Once the dynamical equation for $D^{(\rm tt)}$ is solved, the conformal mode $S$ 
is totally fixed by equation~\eqref{ConstraintS}. One might be worried by
the appearance of the Green's function $G$ in the projector $P^{_{(\parallel)}i}_{\;\;i}$. If the (00)-Einstein equation is depicted 
in the original $h$ variables, all time derivatives drop out immediately and there is no 
doubt that this equation is a constraint. This is no surprise since the theory under consideration is simply 
higher dimensional GR for which this property of the $(00)$--equation is well known. (Compare also to the discussion after equation \eqref{obedience}, where the equivalence to the Hamiltonian constraint in ADM variables is explicitly shown.) The non-local operator is just 
a relict of the graviton decomposition \eqref{deco} and totally fixed
by imposing some arbitrary boundary conditions. Therefore, it cannot spoil the constraint 
character of this equation. The important result is again that $S$ is not an independent 
degree of freedom. Again it should be checked that there is no inconsistency between 
the constraint and the dynamical equation. In total agreement to the GR example the 
solution of the dynamical equation for $D^{(\rm tt)}$ can be inserted in the constraint which yields
\begin{equation}\label{ConstraintS2}
	\Delta_3 S + P^{_{(\parallel)}i}_{\;\;i}\Delta_n S
	=  
	-\frac{2}{3} \frac{n-1}{n+2}\; \kappa_n \; P^{_{(\parallel)}i}_{\;\;i} \; \widehat{t}_\gamma^{\; \gamma}
	\;.
\end{equation} 
It is consistent with \eqref{eomss1} in a sense that 
every solution of~\eqref{ConstraintS2} is a solution of 
\eqref{eomss1} but not the other way around.  

The spectrum of observable quanta is reduced to the propagating degrees of freedom
carried by $D^{(\rm tt)}$, i.e.~$\left[D^{(\rm tt)}\right]\le 5$, due to the index
symmetry and four transversality and one traceless condition.  
For $j\in\{1,2,3\}$, the $(0j)$--Einstein equations are constraints, eliminating
further three components of $D^{(\rm tt)}$ from the list of propagating degrees
of freedom. Since all the gauge redundancies are fixed, all geometrical conditions
exploited, and all constraints imposed, it follows that the localized source 
(\ref{locsour}) absorbs and emits $\left[D^{(\rm tt)}\right] = 2$ physical 
degrees of freedom, subject to a healthy dynamics~(\ref{eomss2}).
  
The localized source (\ref{locsour}) could be an external graviton absorber and
emitter, or it could be resolved in dynamical degrees of freedom.
In fact, both source types could be operational. External sources describe 
those absorbers and emitters that are absolutely inert against backreaction.
They are not the result of integrating out dynamical fields qualifying as heavy 
relative to a preset finite cut--off scale. Integrating out heavy fields on 
$\mathcal{M}_{_4}\times \{y_0\}$ results in an additional Einstein--Hilbert term
localized at $y_0$:
\begin{equation}
	\label{effsour}
	T_{AB}
	=
	L_{AB}^{\; \; \; \; \; \alpha\beta}(y_0,y) \left(t_{\alpha\beta}
	+ \lambda \; G^{(4)}_{\; \; \alpha\beta}(h)\right)(x,y)
	\; .
\end{equation}
Here, $G^{(4)}$ denotes the four--dimensional perturbed Einstein tensor linear in $h$.
Note that action \eqref{ActionGRSource} together with \eqref{effsour} is exactly the linearized BIG model in n co-dimensions \eqref{model}. 
The coefficient $\lambda$ depends on the details of the heavy field theory. For 
phenomenological reasons, $\lambda=1/\kappa_0$ is an attractive choice. 
In this language, $t$ contains, in principal, both, external and dynamically resolved
graviton absorbers and emitters.

The source modification (\ref{effsour}) allows for a straightforward generalization
of (\ref{eomss1}) and (\ref{eomss2}):
\begin{eqnarray}
	\label{eomss3}
	&&\Box S
	=
	\frac{2}{3}\frac{n-1}{n+2}\; \kappa_n
	\left(-\widehat{t}_\mu^{\; \mu} + 3 \; \kappa_0^{-1} \; \Box_4 \widehat{S}
	\right)
	\; , \\ \nonumber \\
	\label{eomss4}
	&& \Box D^{(\rm tt)}_{\;\;\;\;\alpha\beta}
	=
	\kappa_n \left(- 2\; \widehat{t}^{(\rm tt)}_{\;\;\;\;\alpha\beta}
	-\kappa_0^{-1} \; \Box_4 \widehat{D}^{(\rm tt)}_{\;\;\;\;\alpha\beta}\right)
	\; ,
\end{eqnarray}
where $\widehat{S}\equiv S \delta^{(n)}(y-y_0)$ and 
$\widehat{D}^{(\rm tt)}\equiv D^{(\rm tt)}\delta^{(n)}(y-y_0)$
denote the localizations of
the gauge invariant scalar and the transverse and traceless tensor 
on $\mathcal{M}_4\times\{y_0\}$, respectively.

Remarkably, while $\widehat{D}^{(\rm tt)}$ gravitates like an ordinary
energy--momentum source, $\widehat{S}$ does not. In fact, the localized
gauge invariant scalar seems to weaken its own source, which is an
inconsistent modification of the equivalence principle at the classical level 
and indicative for a strong violation of unitarity. 
Assuming the gauge invariant scalar and its localized cousin are dynamical,
the propagator corresponding to (\ref{eomss3}) for the conformal mode~$S$
would exhibit a tachyon pole with a wrong sign residue. 

This is precisely 
the result of \cite{Dubovsky:2002jm,Hassan:2010ys}.
The analysis performed in \cite{Dubovsky:2002jm}
uses a slightly different deconstruction of the graviton, and, as a consequence,
the unitarity violation was claimed to be communicated by $h_\alpha^{\; \alpha}$. 
This can be easily mapped to the deconstruction (\ref{deco}) since
$h_\alpha^{\; \alpha}=B+4 S$, where $S$ is gauge invariant, while $B$
is gauge fixed. Hence, choosing as a gauge condition $B=0$, it is straightforward to
show that (\ref{eomss3}) agrees with the corresponding equations in former treatments.  

However, as before, the $(00)$--Einstein equation gives a constraint on $S$
\begin{multline}\label{obedience}
\left[\partial^i\partial^j\!-\delta^{ij}(\Delta_3\!+\!\Delta_n)\right] D_{\;\;\;ij}^{(\rm tt)}\! +\!\frac{n+\!2}{n-\!1}\left[\Delta_nP^{_{(\parallel)}i}_{\;\;i}\!\! +\!\Delta_3\right]\!S\\
=\kappa_n \left\{ 2 \widehat t_{00} +\kappa_0^{-1}\left(2\Delta_3 \widehat S-\left(\partial^i\partial^j\!-\delta^{ij}\Delta_3\!\right) \widehat D_{\;\;\;ij}^{(tt)}\! \right) \right\}\;,
\end{multline}
which agrees with (\ref{ConstraintS}) in the limit when the heavy fields
on $\mathcal{M}_4\times\{y_0\}$ are formally decoupled from gravity. 
Thus, exactly the same reasoning as before applies.	
As an important result, the effective source (\ref{effsour}) 
does not absorb or emit propagating scalars, which therefore will no challenge its unitarity. 
Equation \eqref{obedience} is equivalent to the Hamiltonian constraint $\Phi_3^{(2)}$ in~\eqref{HamiltonianConstraint}.
This can be checked by explicitly translating the covariant variables to the ADM variables of Section~\ref{hamilton_analysis}. 
From \eqref{deco} together with \eqref{ADM1} and \eqref{LinearizedADM} as well as \eqref{S_s} and \eqref{B_S} it follows
\begin{align}
	h_{ij}	&=D_{\,\,ij}^{(\rm tt)} + P^{_{(\parallel)}}_{\;\;ij} B+\delta_{ij}S \notag\\
		&=D_{\,\,ij}^{(\rm tt)} - 4 P^{_{(\parallel)}}_{\;\;ij} S+\delta_{ij}S\;, \label{ADMtoCov1}\\
		 \notag\\
	N&=\frac{1}{2}s =-\frac{3}{2}S\;(=L)\,,\\
	\notag\\
	L_5&=0\,,\\
	\notag\\
	N_i&=G^{(\rm v,v)}_{\; \; \; \; i 5} + \partial_5 G^{(\rm v,s)}_{\;\;\;\;i}+ \partial_i F^{(\rm s,v)}_{\;\;\;\; 5}\,, \\
	\notag\\
	L_i&=G^{(\rm v,v)}_{\; \; \; \; i 6} + \partial_6 G^{(\rm v,s)}_{\;\;\;\;i}+ \partial_i F^{(\rm s,v)}_{\;\;\;\; 6}\,,
\end{align}
where it has been used that the two dimensional transverse and traceless tensor vanishes. Inserting these relation in \eqref{HamiltonianConstraint} yields \eqref{obedience} with $\widehat t_{00}=0$ and $n=2$ as expected. 
From \eqref{ADMtoCov1} is becomes clear that the non-local term in the constraint is just a relict of the decomposition.
As before it can be checked that the constraint~\eqref{obedience} is in accordance with the would-be dynamical equation~\eqref{eomss3}.

It should be stressed that this analysis is independent of the brane regularization. 
Substituting the delta function with some finite width profile function will
not change the results and conclusions of this section. The $(00)$--Einstein equation
still is a constraint equation even if the source is allowed to have some spread in the extra space directions. The delta function was only taken for the sake of convenience. (This issue
is more difficult to handle if one is interested in the brane-to-brane propagator. In order to regularize its divergencies, one normally has to introduce a certain thickness of the brane to which the final expression for the propagator is sensitive in an essential way~\cite{Hassan:2010ys}. The crossover properties of the theory
for example strongly depend on this choice and setting the brane width to zero corresponds 
to an infinite crossover length scale which would mean that GR would not be modified.  A typical 
choice is the inverse bulk scale $M_{4+n}=\kappa^{-1/(2+n)}_n$ at which the effective 
field theory description of higher dimensional gravity breaks down.)

In accordance with the GR case in Section \ref{cfpGR}, this result questions the validity of former arguments claiming that the $S$ mode is a ghost. These arguments shall be briefly reviewed.
The diagnostic tool that is normally employed to highlight a ghostly absorption or emission 
process on $\mathcal{M}_4$ is the classical brane--to--brane propagator following from (\ref{eomss3})
and (\ref{eomss4}), or equivalently the source--to--source amplitude 
\begin{align} \label{amplitude}
	\int_{\mathcal{M}_{_4}}\!\!\!\! d^4x \, h_{\alpha \beta} t^{\alpha \beta} = 
	\int_{\mathcal{M}_{_4}}\!\!\!\! d^4p \, t^{\alpha \beta}(p) G_{\alpha \beta \gamma \delta}(p^2) t^{\gamma \delta}(-p)\;,
\end{align}
where the propagator in Fourier space is given by
\begin{align}
	G_{\alpha \beta \gamma \delta}(p^2) = 
	\, & \left( \eta_{\alpha \gamma} \eta_{\beta \delta} - \frac{1}{3} \eta_{\alpha \beta} \eta_{\gamma \delta}  \right) G^{(\rm D)}(p^2)  \nonumber 
	\\
	&+  \eta_{\alpha \beta} \eta_{\gamma \delta} \, G^{(\rm S)}(p^2)
\end{align}
with
\begin{align}
	G^{(\rm D)}(p^2) &=
	 \frac{2}{\kappa_n^{-1} g_n^{-1}(p^2) + \kappa_0^{-1} p^2} \label{greenD} 
	 \;, \\ \notag
	 \\
	G^{(\rm S)}(p^2) &= 
	\frac{2}{\left( \frac{n+2}{n-1} \right) \kappa_n^{-1} g_n^{-1}(p^2) - 2 \kappa_0^{-1} p^2} \label{greenS}
	\;.
\end{align}
The function $g_n(p^2)$ is a solution to the equation
\begin{equation}
	\left(p^2 - \Delta_n \right)g_n(p^2,y) = \delta^{(n)}(y) \label{eqgn}
\end{equation}
evaluated at the brane position $y=y_0=0$: $g_n(p^2) \equiv g_n(p^2,0) $. 
The Green's function $G^{(\rm D)}$ follows from (\ref{eomss3}) and $G^{(S)}$ follows from (\ref{eomss4}), where the partially Fourier transformed ansatz 
\begin{align}
	G^{(\rm D,S)}(x,y)=\int d^4p \, e^{ipx}g_n(p^2,y)f^{(D,S)}(p)
\end{align}
has been used.


It is well known that, given phenomenological interesting choices for the parameters of the theory, 
the denominator of (\ref{greenS}) exhibits a pole with negative sign residue. For example, in two 
co-dimensions
\begin{align}
	g_n(p^2) \propto \ln{\left(1+\frac{\kappa_2^{-1/2}}{p^2}\right)}
	\;,
\end{align}
so that the denominator of (\ref{greenS}) has a negative sign residue for $\kappa_0^{-1} \gg \kappa_2^{-1/2}$.
To solve (\ref{eqgn}), one has to regulate an indefinite momentum space integral. Here, a cutoff $\kappa_2^{-1/2}$ was introduced, 
which equals the cutoff of the effective field theory (\ref{model}). Applying the usual Feynman prescription to this pole 
would lead to the conclusion that the corresponding quantum mechanical amplitude contains a negative imaginary part 
and thus violates unitarity due to the optical theorem. However, this calculation does not take into 
account the fact that $S$ is not dynamical. On the contrary, considering the spectral density of the amplitude of \cite{Hassan:2010ys} (equivalently, equation (\ref{amplitude})) actually suggests that a ghost-like scalar degree of freedom is propagating. 

To guarantee that only physical degrees of freedom are generated from the vacuum, one would need to incorporate the constraints properly. 
This is yet an open task. Therefore, in this work a different and more solid way was chosen by performing an hamiltonian analysis.

\section{Summary}\label{Summary}
In this article, the consistency of gravity induced on a higher co--dimensional surface (\ref{model})
is investigated. These models are of great phenomenological interest and might serve as a 
faithful anchor for technically natural approaches to the challenge posed by the Universe's observed
accelerated expansion. 
The prospects of these models have been threatened by claims~\cite{Dubovsky:2002jm, Hassan:2010ys}
questioning their quantum mechanical stability due to seemingly unitarity violating 
absorption and emission processes of a particular scalar degree of freedom.  

However, the action (\ref{model}) can be derived as an effective low energy description of a stable 
parent theory at higher energies. The heavy degrees of freedom belonging to the parent theory
leave fingerprints in the effective theory (\ref{model}) in terms of an induced Einstein--Hilbert term.
Thus, assuming the heavy degrees of freedom constitute legitimate graviton absorber and emitter sources in accordance with Lorentz invariance and gauge invariance, in particular, there is no physical understanding
for the presence of a ghost like excitation in the theory's spectrum. 

Accordingly, in this article it has been shown that gravity induced on a surface
of arbitrary co--dimension respects unitarity.

As a solid diagnostic tool the classical Hamiltonian on the constraint surface has been 
derived within a full--fledged canonical constraint analysis. Its positive definiteness 
clearly proves that the theory (\ref{model}) is healthy because any ghost--like excitation 
would necessarily result in a classical instability.

This result on its own causes a tension with former results in \cite{Dubovsky:2002jm, Hassan:2010ys}. Thus, in Chapter \ref{Source} a covariant language was used in order to make contact to these works. It has been shown that the $(00)$--Einstein equation is a constraint on the dangerous scalar mode $S$ rendering it non-dynamical. 
Therefore, the mode $S$ cannot be excited as an independent degree of freedom. This fact has not been taken into account properly in \cite{Dubovsky:2002jm, Hassan:2010ys}, since both works indicate that a ghost-like scalar degree of freedom can be found amongst the physical particle content of the theory.




These results open an exciting window of opportunity to consistently 
deform gravity at the largest observable distances. We leave the question
of phenomenological viability for future work, where we intend to confront
the deformation (\ref{model}) with data from supernova observation campaigns
as a first step \cite{Berkhahn:appearsoon}, which already shows the richness the effective theory (\ref{model}) has to offer. 

\section*{Acknowledgements}
The authors would like to thank C\'{e}dric Deffayet, Gia Dvali, Fawad Hassan, Valery Rubakov and Robert Schneider for inspiring discussions. The work of SH was supported by the DFG cluster of excellence 'Origin and Structure of the Universe' and by TRR 33 'The Dark Universe'. The work of FB was supported by TRR 33 'The Dark Universe'. The work of FN was supported by the DFG cluster of excellence 'Origin and Structure of the Universe'.

\appendix

\section{}\label{AppH}
Here the full Hamiltonian is presented, which is neither gauge-fixed nor simplified by applying any constraints. It directly follows from substituting the velocities in the linearized version of \eqref{eq_lagrangesplit} with  (\ref{eq_dotNi})--(\ref{eq_dothijy0}) and calculating the Hamiltonian. 
\begin{widetext}
\begin{multline} \label{TotalH}
\frac{\mathcal{H}}{M_6^4}=\frac{1}{M_6^8} \Pi_{(R)ij}\Pi^{ij}_{(R)}+\frac{1}{M_4^2M_6^4} \,\delta^{(2)}_y\, \Pi_{(I)ij}\Pi^{ij}_{(I)} -\frac{1}{4M_6^8}(\Pi_{(R)i}^{\;\;\;\;i})^2-\frac{1}{2M_6^4M_4^2}\,\delta^{(2)}_y\, (\Pi_{(I)i}^{\;\;\;\;i})^2-\frac{1}{4 M_6^8}\Pi_{(R)i}^{\;\;\;\;i}(\Pi_N+\hat{\Pi}_L)\; \\ 
-\frac{1}{8M_6^8}\Pi_N\hat{\Pi}_L+\frac{3}{16M_6^8}(\Pi_N^2 + \hat{\Pi}_L^2)+\frac{1}{2M_6^8}\Pi_5^2+\frac{1}{2M_6^8}(\Pi_i\Pi^i+\hat{\Pi}_i\hat{\Pi}^i)+\frac{2}{M_6^4}\Pi_{ij}n^{j,i}+\Pi_i(n^{i,5}+N_0^{,i})+\hat{\Pi}_i(n^{i,6}+L_0^{,i})\;\\ 
+\Pi_5(L_{0,5}+N_{0,6})-n_{,a}h_{i}^{i,a}-L_{,5}h_{i,5}^i-N_{,6}h_{i,6}^i-2 L_{,i}N^{,i}-2L_{,5}n_{,5}-2N_{,6}n_{,6}-2(N+L)_{,i}n^{,i}+2 N_{0,6}L_{0,5}+2 n^i_{,5}N_{0,i}\;\\ 
+2 n^i_{,6}L_{0,i}+(h_i^i+2n+2L)_{,5}N^i_{,i}+(h_i^i+2n+2N)_{,6}L^i_{,i}-L_{5,i}L^i_{,5}-L_{5,i}N^i_{,6}+\frac{1}{2} L_{5,i}L_{5}^{,i}+ (2n+h_i^i)_{,6}L_{5,5}-N^i_{,6}L_{i,5}\;\\ 
+\frac{1}{4} F_{(N)ij}F_{(N)}^{ij}+\frac{1}{4} F_{(L)ij}F_{(L)}^{ij}+\frac{1}{2}L_{i,5}L^i_{,5}+\frac{1}{2}N_{i,6}N^i_{,6}-h_{ij,5}N^{i,j}-h_{ij,6}L^{i,j}+\frac{1}{4}h_{ij,a}h^{ij,a}-\frac{1}{4}h_{i,a}^ih_{j}^{j,a}\;\\ 
-(1+\frac{M_4^2}{M_6^4}\delta_y^{(2)})(\delta^{1}\!\!\sqrt{-\gamma}\,\delta^{1}\!R^{(3)}+\delta^{2}\!R^{(3)})-(N+L+n)\delta^1\!R^{(3)} -\frac{M_4^2}{M_6^4}\delta_y^{(2)}n\, \delta^1\!R^{(3)}
\end{multline}
\end{widetext}
Where the following definitions are used:
\begin{multline}
\delta^{1}\!\!\sqrt{-\gamma}\,\delta^{1}\!R^{(3)}+\delta^{2}\!R^{(3)}=-\frac{1}{2}h^{ij}_{,i}h_{k,j}^k+\frac{1}{2}h^{jk}_{,j}h^{i}_{k,i}\\
-\frac{1}{4}h_{jk,i}h^{jk,i}+\frac{1}{4}h^j_{j,i}h^{k,i}_{k}\;,
\end{multline}
\begin{align}
\delta^{1}\!R^{(3)}=h^{ij}_{,ij}-h_{k,i}^{k,i}\;,
\end{align}
and
\begin{align}
&F_{(N)}^{ij}=N^{j,i}-N^{i,j} &F_{(L)}^{ij}=L^{j,i}-L^{i,j}\;.
\end{align}

This Hamiltonian is used to calculate the secondary constraints. On the constraint-surface, it reduces to \eqref{FinalH}.

\section{} \label{deconstruction}
In this appendix the details of the graviton deconstruction (\ref{deco}) 
based on the background space--time isomorphism 
$(\mathcal{M}_d,\eta)\cong (\mathcal{M}_4,\eta)\times (\mathbb{R}^n,\delta)$ are presented. 
The deconstruction \eqref{4DDecomp} in the case of four--dimensional GR is the trivial case 
when the parent space time equals $(\mathcal{M}_4,\eta)$. The global Minkowski coordinate system is split accordingly 
into the Cartesian product $X^A=\left(x^\alpha,y^a\right)$ with obvious 
index ranges. The required projectors are 
\begin{eqnarray}
	P^{_{(\parallel)}}_{\;\; \alpha\beta} 
	&:=&
	\partial_\alpha(x) \; \int_{\mathcal{M}_4} {\rm d}^4\tilde{x} \; 
	G(x-\tilde{x}) \; \partial_\beta(\tilde{x}) 
	\; , \nonumber \\
	P^{_{(\perp)}}_{\;\; \alpha\beta} 
	&:=& \eta_{\alpha\beta} \int_{\mathcal{M}_4} {\rm d}^4\tilde{x} \; \delta^{(4)}(x-\tilde{x}) 
	- P^{_{(\parallel)}}_{\;\; \alpha\beta} 
	\; , \nonumber \\
	p^{_{(\parallel)}}_{\;\; ab}
	&:=&
	\partial_a(y) \; \int_{\mathbb{R}^n} {\rm d}^n\tilde{y} \; 
	g(y-\tilde{y}) \; \partial_b(\tilde{y}) 
	\; , \nonumber \\	
	p^{_{(\perp)}}_{\;\; ab} 
	&:=& \delta_{ab} \int_{\mathbb{R}^n} {\rm d}^n\tilde{y} \; \delta^{(n)}(y-\tilde{y}) 
	- p^{_{(\parallel)}}_{\;\; ab}
	\; ,
\end{eqnarray}
where $\partial_A (X)\equiv \partial/\partial X^A$, and $G$ and $g$
are Green's functions corresponding to the four--dimensional wave and $n$--dimensional
Poisson equation, respectively, equipped with arbitrary but fixed boundary conditions. The 
decomposition of the graviton of course works for every choice.
The transverse and traceless projectors are given by the composite operators:
\begin{eqnarray}
	\mathcal{O}^{(4, {\rm tt})\;\;\mu\nu}_{\; \; \; \; \;\;\alpha\beta} 
	&:=&
	P^{_{(\perp)}\gamma}_{\;\;\; \alpha}\; P^{_{(\perp)}\delta}_{\;\;\; \beta}
	- 
	P^{_{(\perp)}}_{\;\; \alpha\beta} \; P^{_{(\perp)}\gamma\delta}_{}/3 
	\; , \\
	\mathcal{O}^{(n, {\rm tt})\;\;cd}_{\; \; \; \; \;\;ab} 
	&:=&
	p^{_{(\perp)}c}_{\;\;\; a}\; p^{_{(\perp)}d}_{\;\;\; b}
	- 
	p^{_{(\perp)}}_{\;\; ab} \; p^{_{(\perp)}cd}_{}/(n-1) 
	\; . \nonumber 
\end{eqnarray}	
The field content of the graviton in the $(\alpha\beta)$--sector is given by
\begin{eqnarray}
	h_{\alpha\beta}
	&=&
	D^{(\rm tt)}_{\;\;\;\; \alpha\beta} + \partial_{(\alpha} V^{_{(\perp)}}_{\;\;\; \beta)} +
	P^{_{(\parallel)}}_{\;\; \alpha\beta} \; B + \eta_{\alpha\beta} \; S
	\; , \nonumber \\ \nonumber \\
	&& D^{(\rm tt)}_{\;\;\;\; \alpha\beta}
	=
	\mathcal{O}^{(4, {\rm tt})\;\;\gamma\delta}_{\; \; \; \; \;\;\alpha\beta} \; h_{\gamma\delta}
	\; ,\nonumber \\
	&&\partial_\alpha V^{_{(\perp)}}_{\;\;\; \beta} 
	=
	P^{_{(\parallel)}\gamma}_{\;\;\; \alpha} \; P^{_{(\perp)}\delta}_{\;\;\; \beta} \; h_{\gamma\delta}
	\; ,\nonumber \\
	&& B
	=
	\left(P^{_{(\parallel)}\gamma\delta}_{}-P^{_{(\perp)}\gamma\delta}_{}/3\right) h_{\gamma\delta}
	\; , \nonumber \\
	&& S
	=
	P^{_{(\perp)}\gamma\delta}_{} \; h_{\gamma\delta}/3 \; .
\end{eqnarray}
In the $(ab)$--sector, the graviton's field content is consequently given by
\begin{eqnarray}
	h_{ab}
	&=&
	d^{(\rm tt)}_{\;\;\;\; ab} + \partial_{(a} v^{_{(\perp)}}_{\;\;\; b)} +
	p^{_{(\parallel)}}_{\;\; ab} \; b + \delta_{ab} \; s
	\; , \nonumber \\ \nonumber \\
	&& d^{(\rm tt)}_{\;\;\;\; ab}
	=
	\mathcal{O}^{(n, {\rm tt})\;\;cd}_{\; \; \; \; \;\;ab} \; h_{cd}
	\; ,\nonumber \\
	&&\partial_a v^{_{(\perp)}}_{\;\;\; b} 
	=
	p^{_{(\parallel)}c}_{\;\;\; a} \; p^{_{(\perp)}d}_{\;\;\; b} \; h_{cd}
	\; ,\nonumber \\
	&& b
	=
	\left(p^{_{(\parallel)}cd}_{}-p^{_{(\perp)}cd}_{}/(n-1)\right) h_{cd}
	\; , \nonumber \\
	&& s
	=
	p^{_{(\perp)}cd}_{} \; h_{cd}/(n-1) \; .
\end{eqnarray}	
Finally, the field content in the $(\alpha b)$--sector is given by
\begin{eqnarray}
	h_{\alpha b}
	&=&
	G^{(\rm v,v)}_{\; \; \; \; \alpha b} + \partial_b G^{(\rm v,s)}_{\;\;\;\; \alpha}
	+ \partial_\alpha F^{(\rm s,v)}_{\;\;\;\; b} + \partial_\alpha \partial_b F^{(\rm ss)}
	\; , \nonumber \\ \nonumber \\ 
	&& G^{(\rm v,v)}_{\; \; \; \; \alpha b}
	=
	P^{_{(\perp)}\gamma}_{\;\;\; \alpha} p^{_{(\perp)}c}_{\;\;\; b} \; h_{\gamma c}
	\; , \nonumber \\
	&& \partial_b G^{(\rm v,s)}_{\;\;\;\; \alpha}
	=
	p^{_{(\parallel)}c}_{\;\;\; b} \; P^{_{(\perp)}\gamma}_{\;\;\; \alpha} \; h_{\gamma c}
	\; , \nonumber \\
	&& \partial_\alpha F^{(\rm s,v)}_{\;\;\;\; b}
	=
	P^{_{(\parallel)}\gamma}_{\;\;\; \alpha} \; p^{_{(\perp)}c}_{\;\;\; b} \; h_{\gamma c}
 	\; , \nonumber \\
	&& \partial_\alpha \partial_b F^{(\rm ss)}
	=
	P^{_{(\parallel)}\gamma}_{\;\;\; \alpha} \; p^{_{(\parallel)}c}_{\;\;\; b}\; h_{\gamma c}
	\; .
\end{eqnarray}	
\\
Introducing a new coordinate system via the following infinitesimal transformations
\begin{eqnarray}
	\delta x^\alpha 
	&=&
	\partial^\alpha(x)\left(\int {\rm d}^n \tilde{y} \;
	g(y-\tilde{y})b\left(x,\tilde{y}\right)/2 
	- F^{(\rm ss)}(x,y)\right)
	\nonumber \\
	&& - V^{_{(\perp)}\alpha}(x,y)
	\; , \nonumber \\
	\delta y^a
	&=&
	\int {\rm d}^n \tilde{y}\; g(y-\tilde{y}) \partial^a(\tilde{y}) b(x,\tilde{y})
	- v^{_{(\perp)}a}(x,y)
	\; ,
	\end{eqnarray}	
results in the gauge fixed graviton deconstruction (\ref{deco}).

\bibliographystyle{unsrt}

\end{document}